\documentclass[referee,a4paper,12pt,traditabstract]{swsc} 

\usepackage{graphicx}
\usepackage{newtxmath} 
\usepackage{caption}
\usepackage{subcaption}
\usepackage{placeins}
\usepackage{epstopdf}
\usepackage[displaymath,mathlines]{lineno}
\usepackage[authoryear,round]{natbib}
\usepackage[backref]{hyperref}
\usepackage{url}
\usepackage{xcolor}
\usepackage{amsmath}
 
\usepackage[normalem]{ulem}
\usepackage{tcolorbox}

\hypersetup{colorlinks=true,citecolor=cyan,urlcolor=cyan,linkcolor=blue}

\begin{document}

   \title{The radiation environment over the African continent at aviation altitudes: First results of the RPiRENA-based dosimeter}

   \titlerunning{Aviation dosimetry over Africa}

   \authorrunning{Mosotho et al.}

   \author{M.G. Mosotho\inst{1}\thanks{Corresponding author: \email{\href{mailto:mosothomoshegodfrey@gmail.com}{mosothomoshegodfrey@gmail.com}}}, R.D. Strauss\inst{1}, S. B{\"o}ttcher\inst{2}, C. Diedericks\inst{1}}


   \institute{Center for Space Research, North-West University,
              Potchefstroom, South Africa
              \email{\href{mailto:mosothomoshegodfrey@gmail.com}{mosothomoshegodfrey@gmail.com}, \href{mailto:dutoit.strauss@nwu.ac.za}{dutoit.strauss@nwu.ac.za}}
         \and 
        Institut f\"ur Experimentelle und Angewandte Physik, Universit\"at Kiel, Kiel, Germany 
         \email{ 
         \href{mailto:boettcher@physik.uni-kiel.de}{boettcher@physik.uni-kiel.de}}
             }
 
  \abstract
   {The radiation environment over the African continent, at aviation altitudes, remains mostly uncharacterized and unregulated. In this paper we present initial measurements made by a newly developed active dosimeter on-board long-haul flights between South Africa and Germany. Based on these initial tests, we believe that this low-cost and open-source dosimeter is suitable for continued operation over the Africa continent and can provide valuable long-term measurements to test dosimteric models and inform aviation policy.}       

\maketitle

\section{Introduction}

Primary cosmic-rays (CRs), in particular those of galactic and solar origin, are the most dominant particle species interacting with atmospheric molecules creating a hazardous secondary radiation field at, and above, aviation altitudes {\citep{Reames1999SolarEP, Vainio_etal2009,Desai_Giacalone2016}}. The secondary radiation produced at aviation altitudes is heavily depended on several factors, including the phase of the solar cycle, the geomagnetic cutoff rigidity ($\rm P_c$) determining which CRs can penetrate the protective magnetosphere, the residual atmosphere about the aircraft, and relevant aircraft shielding \citep{SMARTSHEA2005,Feynman_Gabriel2000,Desmaris_2016}. The International Committee on Radiological Protection (ICRP) has recognized and classified flight personnel's CR exposure as an occupational hazard. The ICRP guidance recommends that exposure to ionising radiation should be below 20 mSv/yr for radiation workers, while 1 mSv/yr is recommended for the general public and occasionally exposed commercial flight passengers \citep{ICRP60,ICRP132}. {{In Europe and USA studies by e.g.}} {\citet{Thierfeldt2009}} {{and the US Federal Aviation Administration}}\footnote{\url{https://www.faa.gov/data_research/research/med_humanfacs/aeromedical/radiobiology}} {{have shown that the annual maximum effective dose for regular passengers and pilots are about 6$-$7 mSv.}} {{In Africa it is unclear what the annual effective dose range for pilots or regular passengers will be at aviation altitudes as no assessment of their radiation exposure has been conducted. Nonetheless, a}}t flight altitudes, radiation exposure during a single flight will normally not cause immediate health effects or exceed any recommended limits. But, for flight {{personnel, exposure}} increases their long-term radiation dose accumulation and this might ultimately contribute to an additional health risk. However, it is possible that the minimum exposure recommendation can be exceeded in a single flight during a severe solar particle event \citep{Matthia_etal2009,Fujita_etal2021}. On the Earth's surface, the temporal flux of cascading particles produced within the atmosphere has been monitored for years using a network of ground-based neutron monitors \citep[NMs, e.g.][]{simpson2000,Moraal_etal2000}. With the recent record-breaking CR fluxes observed from NM stations during the previous solar minima in 2009 and 2020 \citep[e.g.][]{MoraalStoker2010,Oh_etal2013,strausspotgieter2014,Fuetal2021}, there is additional motivation to routinely monitor radiation levels at aviation altitudes even during very quiet solar periods. \\

To support efficient and safe air navigation, the International Civil Aviation Organization's (ICAO's) revised airspace procedural manual includes expected radiation dose exposure as part of the mandatory information to be provided from designated space weather centres as part of flight planning \citep{ICAO2019,ICRP132}. In the African airspace, the South African National Space Agency's (SANSA's) space weather center\footnote{\url{https://spaceweather.sansa.org.za/}} provides advisories to the Air Traffic and Navigation Service (ATNS) regarding enhanced radiation levels at flight altitudes for civil aviation\footnote{\url{https://www.sansa.org.za/2021/10/14/sansa-and-atns-signs-mou/}}. There is thus a need to characterise and quantify the localized radiation levels on-board commercial aircraft in near real-time using active dosimeters \citep[see][]{Ritter_etal2014,Moller_thesis2013}.\\

To meet the requirements for space weather application, a suitable dosimeter must ideally to be lightweight and relatively small to make it easy to deploy in different locations on ground-level, on-board commercial aircraft flights, or even during stratospheric balloon launches. Apart from being a {\it plug and play} instrument, it should also have very low power consumption and have an environmental sensor to monitor (cabin) pressure and temperature. For this work, the RPiRENA\footnote{See a detailed description of the project at \url{http://www.ieap.uni-kiel.de/et/people/stephan/rpirena/}} (i.e. RPi stands for Raspberry Pi and RENA stands for Read-out Electronics for Nuclear Applications) detector, which is {{an open-source project (instrument schematics, all software, and calibration techniques are available publicly)}}, was modified to serve as such an active dosimeter.\\

This instrument and its calibration is discussed in more detail in the appendices. {{The radiation field at flight altitudes differs from that at ground level, as well as that produced by the calibration sources used here. The detector is calibrated for the low energy range of the secondary CR radiation, while, for the higher energies, we assume a linear extrapolation up to 7 MeV in silicon. This was shown to be a reasonable assumption by}} {{\cite{Ritter_thesis2013}}} {{using the same low energy calibration sources}}. In the current configuration, the dosimeter includes two silicon PIN diodes that can be operated in single mode, where we obtain the absorbed dose in each diode, or in telescope mode which allows us to determine the linear energy transfer and therefore also the quality factor of the incident radiation. In this work we present initial results of this newly developed dosimeter during two long-haul flights from South Africa to Germany.\\

\section{Dosimetric quantities}
 
When incident ionizing radiation interacts with matter, they deposit energy in the material. The energy per unit mass deposited in the target material, by the incident ionising radiation, is a quantity called the absorbed dose. For our silicon diodes (SDs), this quantity is obtained from the energy deposition spectra of each SD as

\begin{equation}
\mathcal{D}_{\rm{Si}}=\frac{1}{M} \sum_{i}^{} \mathcal{N}_i \cdot E_i,
\end{equation}

where $\mathcal{N}_i$ is the number of events detected in the $i^{th}$ energy bin, with midpoint located at ${E_i}$, and $M$ is the SD's mass. The energy deposition of each particle (or event) is calculated by using the calibrated conversion described in Sec. \ref{subsec:calibration}. Absorbed doses measured using silicon-based detectors are given in units Gy. To indirectly mimic the biological damage in human tissue as a result of exposure to ionising radiation, the absorbed {{dose}} measured in silicon has to be converted to a comparable value in water, $\mathcal{D}_{\rm{H_2O}}$, with the absorbed dose in water now given in units of Sv \citep[see][]{ICRP103,HellwegChrista2007,Bartlett_2004,Flowers_etal2018}. We use the standard approach for this conversion,

\begin{equation}
    \mathcal{D}_{\rm{H_2O}} = \mathcal{F}_{ \rm{Si} \rightarrow \rm{H_2O}} \cdot D_{\rm{Si}},
\end{equation}

where $\mathcal{F}_{ \rm{Si} \rightarrow \rm{H_2O}}=1.2$ \citep{Beaujean_etal2005}. A corresponding absorbed dose rate, $\mathcal{D}_r$, can also be calculated by dividing $\mathcal{D}_r$ with the time interval of the measurement. With the dosimeter operated in telescope mode, coincidence events (an event detected simultaneously in both diodes) allow us to {{approximate}} the linear energy transfer (LET) spectrum of the incident radiation. The ionizing particle moves through one SD, depositing an amount of energy $\Delta E$, after which it is detected in the second diode. The linear transfer, again in silicon, is then {{approximated}} as

\begin{equation}
    \mathcal{L}_{\rm{Si}} = \frac{dE}{dl} \approx \frac{\Delta E}{\langle l_{\mathrm{eff}} \rangle},
\end{equation}

where $\langle l_{\mathrm{eff}} \rangle$ is the effective thickness of the SD, calculated in Sec. \ref{subsec:thickness}. {{In this approach,}} the LET is therefore {{estimated as}} the energy loss of charged particles interacting with the target material per path length and allows us to {{approximate}} the quality factor of the incident radiation and therefore its biological effectiveness. Similar to the absorbed dose, the LET in silicon needs to be converted to LET in water. This is done through the expression
	
\begin{equation}
	\label{eq:l_h20}
	{\mathcal{L}_{\rm{H_20}}} \approx \frac{\Delta E}{ \langle {l}_{\mathrm{eff}} \rangle } \cdot \mathcal{F}_{ \rm{Si} \rightarrow \rm{H_2O}} \cdot \frac{\rho_{\rm{H_2O}}}{\rho_{\rm{Si}}},
\end{equation}
	
where $\rho_{\rm{H_2O}}$ and $\rho_{\rm{si}}$ are the densities of water and silicon with values of 1 g/cm$^3$ and 2.33 g/cm$^3$, respectively \citep[see][]{Ritter_etal2014} and, once again, $\mathcal{F}_{ \rm{Si} \rightarrow \rm{H_2O}}=1.2$. 

\section{First Results}
\label{Sec:Results}

\begin{figure}[ht]
		\centering
	\includegraphics[width=\textwidth]{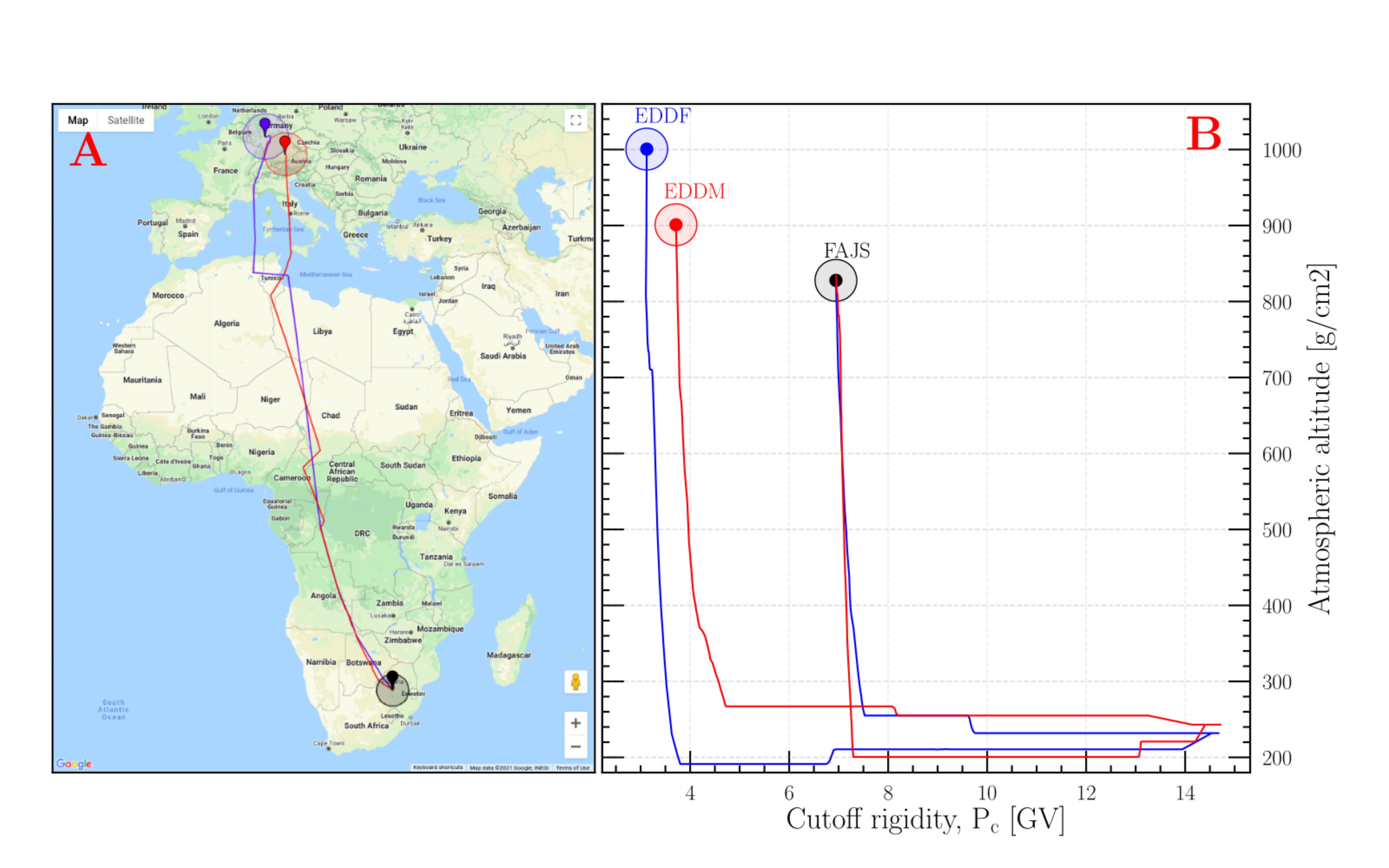}	
	{\caption[]{\label{Fig:GoogleMap}{\textnormal{Panel A shows the test flight routes from FAJS to EDDF (in blue) while the return flight to FAJS from EDDM is shown in red. Panel B shows the flight altitudes (in terms of atmospheric depth) as a function of the corresponding vertical geomagnetic cutoff rigidity, $\rm P_c$.}}}}
\end{figure}

\begin{figure}[ht]
		\centering
	\includegraphics[width=\textwidth]{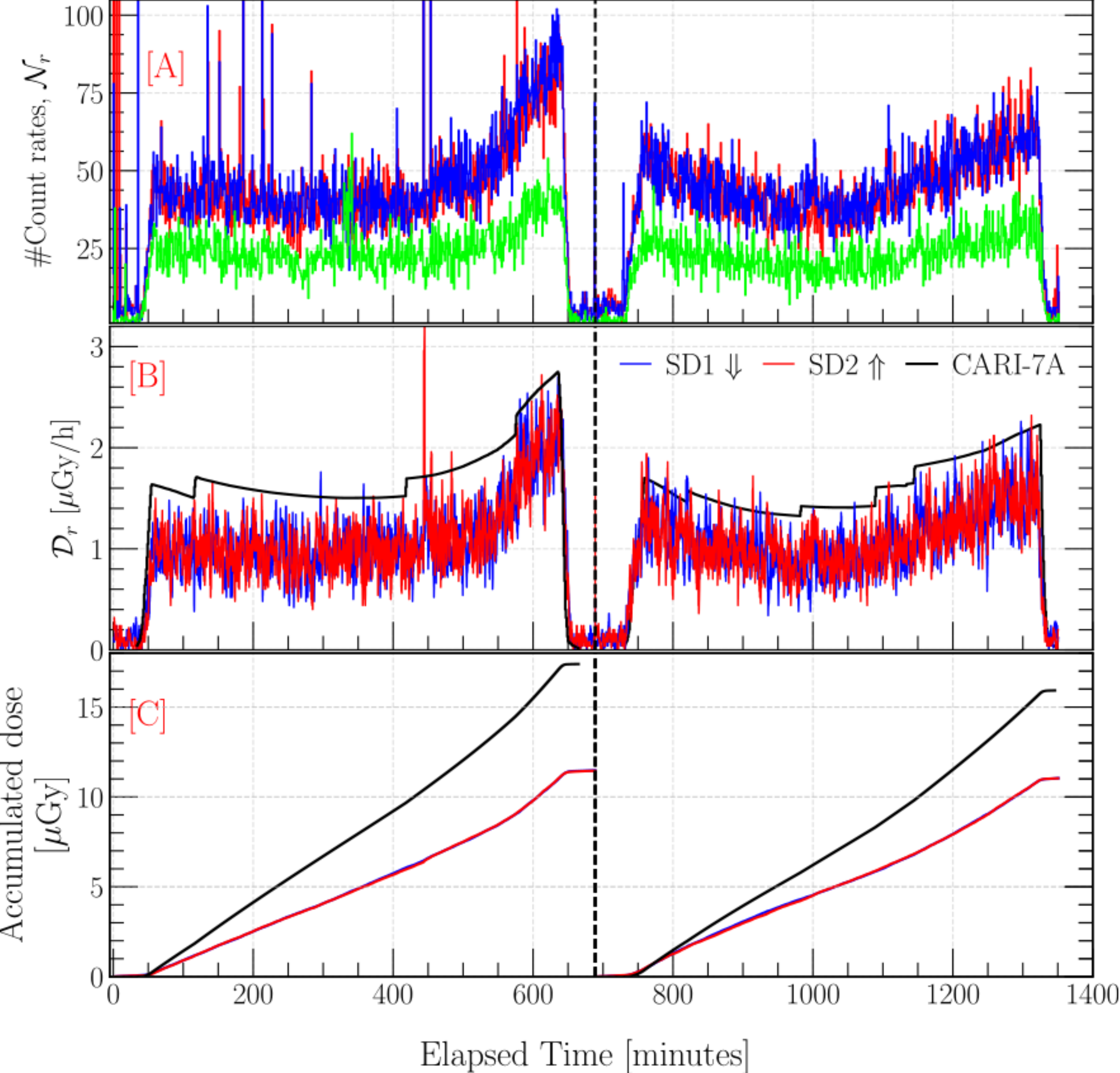}	
	{\caption[]{\label{Fig:DoseRates}{\textnormal{Panel A shows the measured count rates (SD 1 in blue, SD 2 in red, and coincidence counts in green) during the two tests. Panel B shows the measured  absorbed dose rate in each diode, with a minute cadence, while panel C shows the accumulated absorbed dose, {{$\mathcal{D}_{a}$}}. The black curves in the bottom panels show results for the CARI-7A model {{for 0.3 mm of silicon}}. The left panels are for the flight from FAJS to EDDF and the right panels for from EDDM to FASJ.}}}}
\end{figure}

\begin{figure}[ht] 
	\centering
	\includegraphics[trim=0mm 0mm 0mm 0mm,clip,scale=0.5]{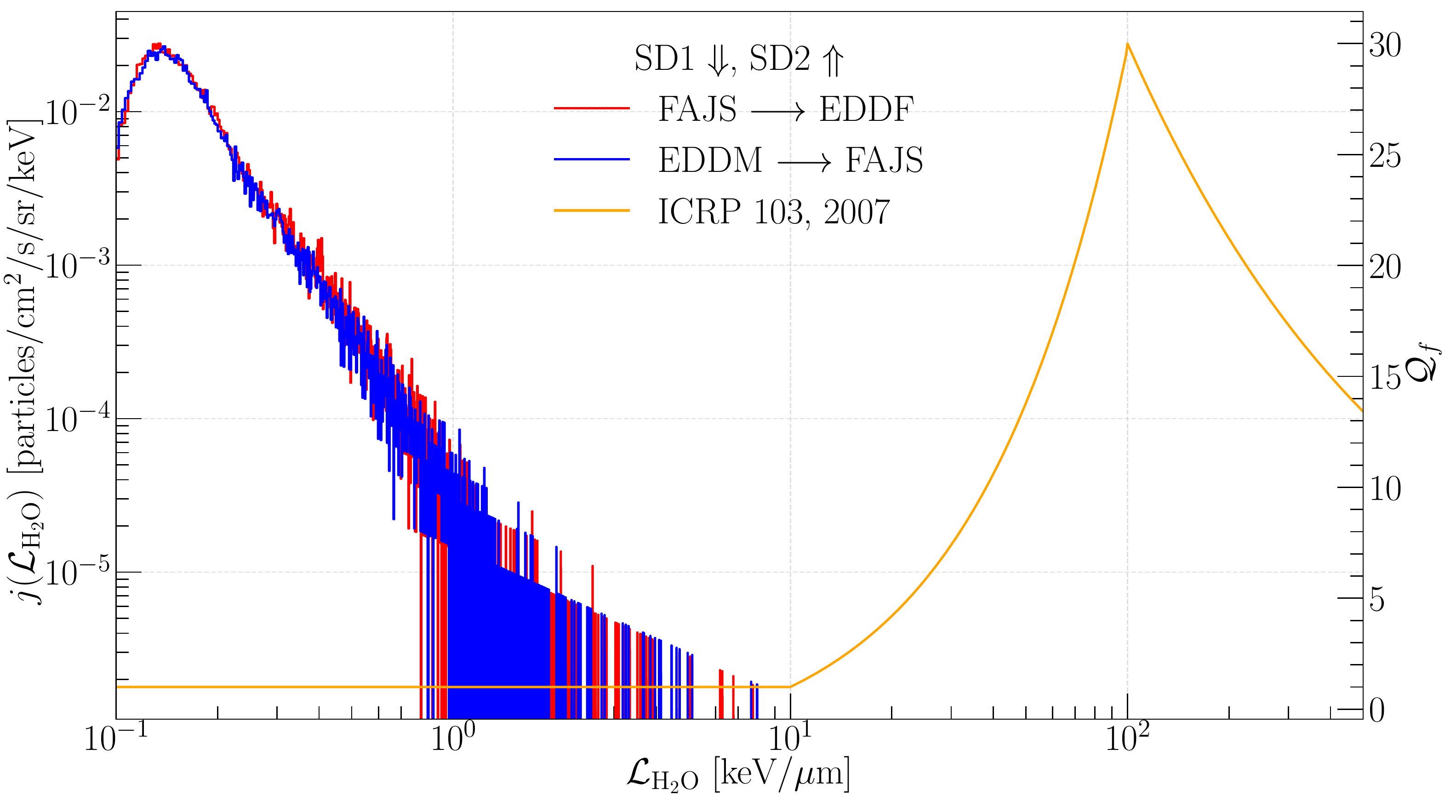}
	{\caption[]{\label{Fig:EnerSpecGERM_SA}{\textnormal{LET spectra accumulated for the flight to EDDF (in red) and a return flight to FAJS (in blue). The quality factor, as defined by \citet{ICRP103}, is shown in orange.  }}}}
\end{figure}

The dosimeter was placed on two long-haul flights: 10 June 2019 from Johannesburg (FAJS) to Frankfurt (EDDF) and on 15 June 2019 from Munich (EDDM) to FAJS, each lasting between 11 and 12 hours. These measurements were performed inside a commercial aircraft and serve as a first test of the instrument in a mixed radiation field over the African continent. To interpret radiation dose rate measurements obtained onboard aircraft, it is essential to discuss the possible space weather effects and influences on the measurements. During these flights, the geomagnetic conditions were assessed using the Kp index and in both flights it was below 3, an indication that the magnetosphere was virtually undisturbed\footnote{See data from omniweb: \url{https://omniweb.gsfc.nasa.gov/form/dx1.html}}. {{Furthermore, there were no solar events during these flights. Thus, the contribution}} of solar energetic particles can be excluded. The omnipresent GCR component observed from the NMDB\footnote{The Neutron Monitor Database (NMDB): \url{http://www.nmdb.eu/}} polar stations show marginal modulation variation rates below 2\% for both flight routes. As a result, changing space weather effects during both flights can be assumed to be negligible.\\

The left panel of Figure~\ref{Fig:GoogleMap} shows a visualization of the flight routes, while the right panel shows the corresponding altitude and effective vertical cut-off rigidity, $\rm P_c$, during these flights. {Data for the flight routes data were downloaded from FlightAware\footnote{\url{https://flightaware.com/}}}. For these long-haul flights, the $\rm P_c$ varied significantly, covering both the North and Southern hemispheres, and crossing the magnetic equator, during each flight. The relevant flight data for flights are summarized in Table~\ref{tab:SDsetUp}. \\
 
 \begin{table}
	\centering
	\caption[]{Long-haul flight information.}
	\label{tab:SDsetUp}
	$$ 
	\begin{array}{p{0.009\linewidth}lllllll}
	\hline	\hline
	\noalign{\smallskip}
	\centering
	\textbf{SDs}     &~~~ &   ~~~	\textbf{Route}	& 	~~	\textbf{Date}		&\textbf{Duration} & \textbf{Cruise Altitude} & ~~~~~~~~~~~~\textbf{{$\mathcal{D}_{a}$}} \\
	& & ~~   &  & ~~~~~\textbf{[hr]} & \textbf{~~~~~~~[g/cm$^{2}$]} & ~~~~~~~~~~\textbf{[$\mu$Gy]}\\
	\noalign{\smallskip}
	\hline
	\noalign{\smallskip}
	$\Downarrow\Uparrow$ & &     ${\rm FAJS - EDDF}$    &  ${10/06/2019}$  &  ~~~~$11.43$ &   $ 255.05-191.29 (4){$^{\rm a}$} $   &  $\rm{11.48 (SD 1)~\&~11.46 (SD 2) }$ \\
	
	$\Downarrow\Uparrow$  & &    ${\rm EDDM - FAJS}$    &  ${15/06/2019}$  &  ~~~~$10.96$ &  $ 267.10-200.56 (5){$^{\rm a}$} $   &  $\rm{11.04 (SD 1)~\&~11.03 (SD 2) }$ \\
	\noalign{\smallskip}
	\hline
	\end{array}
	$$ 
	\begin{flushleft} 
		$^{\mathrm{a}}$ \scriptsize{Low and upper limit of cruise altitude, while the numbers in brackets indicates the number of times the cruise altitude has changed, this includes the first cruise altitude}.\\	
	\end{flushleft}
\end{table}

In Figure~\ref{Fig:DoseRates} we show the count rates from both SDs during both test flights. The panels on the left {{represent}} the flight from FAJS to EDDF and the panels on the right represent the flight from EDDM to FAJS. Blue and red curves show the results for both diodes, while the green curves is the count rates of the coincident events. In panel A, during take-off and landing, it can be seen that with increasing altitude (decreased atmospheric shielding) the number of detected particles increases sharply. During the cruising phase the count rate changes due to changing values of $\rm P_c$ (note the decrease towards the equatorial regions) and changes in the flight altitude, especially the noticeable increase towards the end of the first flight. \\

In panel B, the dependence of the calculated absorbed dose rate in silicon on the altitude and $\rm P_c$ is also visible. At these altitudes ionizing particles, e.g. muons, electrons, and protons, make up the bulk of the particles contributing to the absorbed dose rate measured in silicon. In panel C, {{an accumulated}} absorbed dose {{($\mathcal{D}_{a}$)}} of $\sim$11.5 $\mu$Gy was recorded on the flight from FAJS to EDDF in both diodes, while a dose of $\sim$11 $\mu$Gy was recorded on the return flight. The data for the two flights in terms of {{an accumulated}} flight dose are also presented in Table~\ref{tab:SDsetUp}. The measured absorbed doses were compared to CARI-7A\footnote{\url{https://www.faa.gov/data_research/research/med_humanfacs/aeromedical/radiobiology/cari7}} model simulations{{, for model information see e.g. \citet{Copeland2017}}} {{while an overview of the model calculations performed (and shown in panel B and C) are given in Appendix~\ref{Append:B}. For both flights, the model predicts higher accumulated absorbed doses, i.e. about a factor of $\sim$1.5 higher, compared to the measurements.}}\\


The generated LET spectra for both flights, integrated over the total flight duration, are given in Figure~\ref{Fig:EnerSpecGERM_SA}. The spectrum is constructed by making use of the calculated geometrical factor of the SDs in telescope mode as discussed in Sec. \ref{subsec:geometrical}. The LET spectrum of the FAJS to EDDF flight is given in red and the return flight is in blue. Clearly visible is the prominent peak from minimal ionizing charged particles and the steep decrease towards higher energy depositions following a power law. Figure~\ref{Fig:EnerSpecGERM_SA} also shows, in orange, the quality factor from \citet{ICRP103}. We can combine the LET spectra, and the dependence of the quality factor on the LET, to determine the mean quality factor as

\begin{equation}
\label{Eq:Qf}
    \langle \mathcal{Q}_f \rangle = \frac{\int j(\mathcal{L}_{\rm{H_20}}) \mathcal{Q}_f (\mathcal{L}_{\rm{H_20}}) d\mathcal{L}_{\rm{H_20}}}{\int j(\mathcal{L}_{\rm{H_20}}) d\mathcal{L}_{\rm{H_20}}}.
\end{equation}

As we have no significant LET events with $\mathcal{L}_{\rm{H_20}} > 10$ keV/$\mu$m, we find  $\mathcal{Q}_f=1$. The calculated mean quality factor does, however, deviate strongly for individual flights since it depends strongly on the few high LET events that are observed during the rather short measurement intervals with a dosimeter with a small detection area {\citep[see][]{Ritter_thesis2013,Ritter_etal2014}}. Using the calculated $\langle \mathcal{Q}_f \rangle$, the dose equivalent (in water) can be  calculated from the calculated absorbed dose in water as
\begin{equation}
\label{Eq:H}
    \mathcal{H}_{\rm{H_2O}} = \mathcal{D}_{\rm{H_2O}} \cdot \langle \mathcal{Q}_f \rangle.
\end{equation}
 
\section{Discussion and Conclusion}

Much research has been devoted to studying the effects and possible impacts of radiation exposure during commercial flights. ICAO's recent amendments specifically focus on the short-term impacts and effects of radiation exposure on avionics and the health and safety of flight personnel. As such, near real-time and reliable estimates of the radiation levels at aviation altitudes is essential. In this manuscript, an assembled, tested, and calibrated dosimeter, based on the open-source RPiRENA detector, is described. Initial long-haul test flight of this prototype are very promising and present the first attempt at characterizing the radiation levels at aviation altitudes over the African continent.\\

We believe that the RPiRENA-based dosimeter described in this work offers the opportunity to expand pre-existing coverage of the existing dosimeter network, especially over Africa. Measurements from this instrument are, and will continue to be, of high importance to establish, enrich, and maintain a dialogue regarding radiation dosimetry and radiation protection between general members of the public, academia, and the private sector, especially within African airspace, where there is little or no data available. \\

Since the prototype discussed in this manuscript was tested, the instrument has been developed further, including the addition of a GPS antenna to directly log the position (when such a measurement is possible) of the instrument and refining the software to be more user-friendly. One drawback of the current design, and this is in general true for all active dosimeters using SDs, is the inability to correctly account for the absorbed dose due to neutrons, especially at aviation altitudes. {{Due to the SD-based detector's small interaction cross section with neutrons \citep[][]{Ritter_thesis2013,Ritter_etal2014}, the SDs used in this work have a low efficiency for neutron detection. At cruising altitudes it is believed that the radiation absorbed dose is dominated by the absorbed dose received from neutrons, generated as a result of the interactions of primary CRs with atmospheric molecules \citep[][]{Bartlett_etal2002,Poje_etal2015,Meier_etal2020} and subsequently with the aircraft components \citep[][]{Knipp_2017}.}} Work is, however, underway locally in South Africa to account for the neutral component at flight altitudes through the development of a dedicated neutron detector \citep{buffleretal2015} that can be flown and operated independently alongside the silicon-based dosimeter discussed here in order to measure the dominant components of the complex radiation field at aviation altitudes.\\

\acknowledgements

The author acknowledges financial support from the South African National Space Agency's (SANSA, Space Science Division) PhD bursary program and the National Research Foundation (NRF) of South Africa's Professional Development Programme (PDP). The  NRF's  unique identifier number is 104898 -- S$\&$F -- Scholarships $\&$ Fellowships Programme. Opinions expressed and conclusions arrived at are those of the authors and are not necessarily to be attributed to SANSA or the NRF. We acknowledge support provided by the North-West University (NWU) Instrument Making Department. Figures in this paper were prepared with Matplotlib \citep{Hunter_2007}.

\clearpage

\appendix

\section{The RPiRENA-based active dosimeter}

\subsection{The detector's working principle and signal processing}

\begin{figure}[ht]
		\centering
	\includegraphics[width=\textwidth]{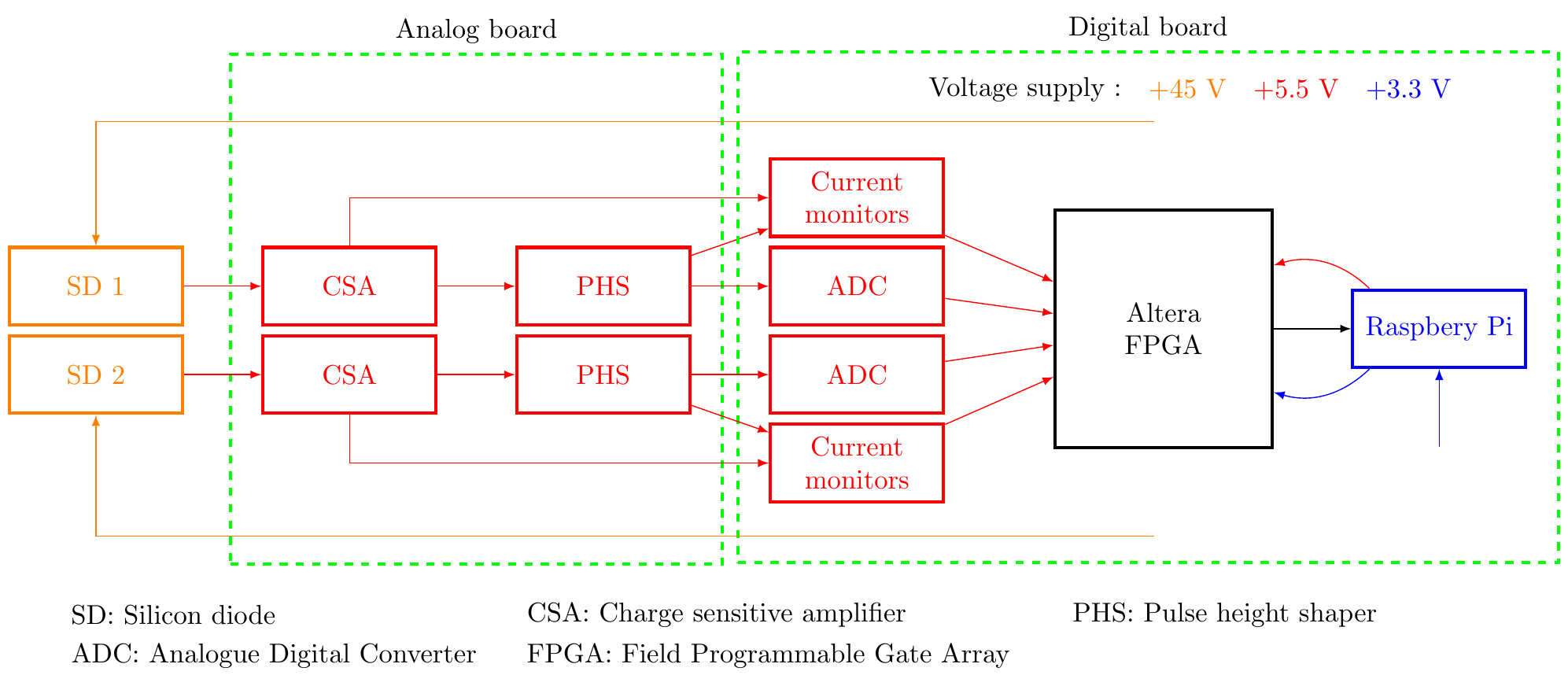}
	\caption{\label{Fig:Block_diagram2}{\textnormal{Schematic representation of the RPiRENA detector, divided into an analog and a digital part.}}}
\end{figure}

Figure~\ref{Fig:Block_diagram2} shows a schematic representation of RPiRENA detector. This particle detector works on the same principles as previously assembled dosimeter prototypes e.g. the Mobile Dosimetric Telescope \citep{Ritter_etal2014,Ritter_thesis2013}. Here, the detection system draws its power from the 3.3 V and 5 V GPIO (general purpose input output) pins from the connected Raspberry Pi (RPi) zero micro-computer, which is in turn powered via a rechargeable power bank. The instrument draws about 0.4 A of power. Data processing and storage is also performed on the RPi.\\

The silicone diodes (SDs) are connected to a reversed bias voltage of 45 V to fully deplete the 300 $\mu$m thick diodes. When ionising radiation interacts with a SD a short charge pulse is generated due to the separated charges \citep{Leo_1987}. However, the charge pulse is too small to be sensed directly. Therefore, the pulse is first sent to the amplifier section of the analogue electronics. Here, a charge-sensitive amplifier (CSA) amplifies the signals before the resulting pulse is shaped (by the pulse-height shaper, PHS), then it is digitized by a 16 bit analog-to-digital converter (ADC). A field programmable gate array (FPGA) is programmed to periodically sample the ADC signals and detect the peak voltage (i.e. the pulse height). Since the pulse height is proportional to the energy deposited, the ADC value is proportional to the energy deposited in the detector after the appropriate calibrations are performed.\\

\subsection{The prototype detector setup}

\begin{figure}[ht]
    \centering
    \includegraphics[width=0.3\textwidth]{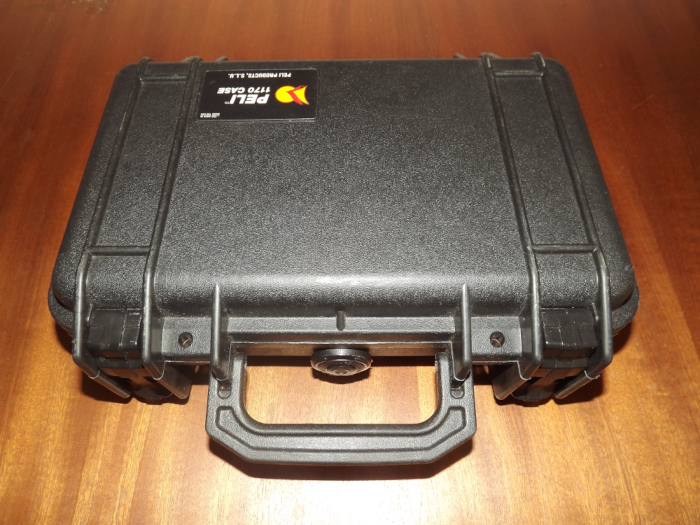}
    \includegraphics[width=0.3\textwidth]{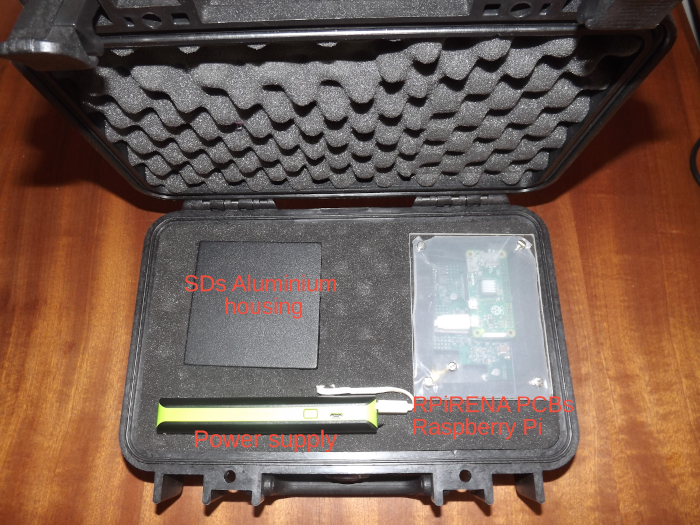}
    \includegraphics[ width=0.38\textwidth]{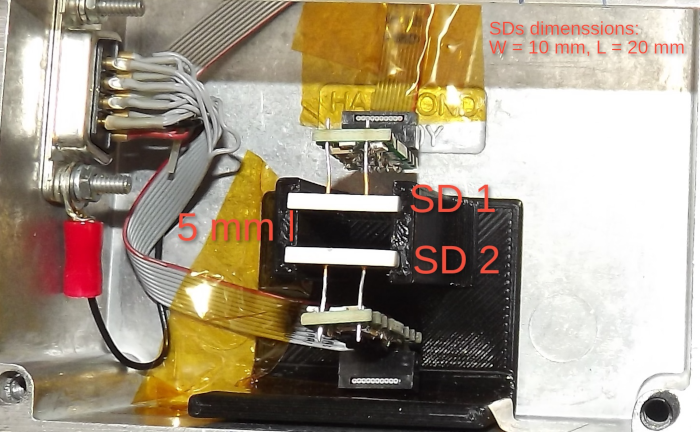}
    \caption{External (left panel) and interior (middle panel) dosimeter housing used for the test flights. The right panel shows a close-up view of silicone diodes (SDs), labelled SD 1 and SD 2, in telescope mode with a separation of 5 mm.}
    \label{Fig:Housing}
\end{figure}

Figure~\ref{Fig:Housing} shows an exterior and interior view of the first RPiRENA-based prototype assembled and tested in 2017. The prototype is placed in a plastic protective housing (shown in the left panel), while the two electronics boards (i.e. the FPGA circuitry and the RPi) along with a power bank and the light proof SD housing are shown in the middle panel. The right panel shows the two SDs in telescope mode. 

\subsection{Calibration}
\label{subsec:calibration}

\begin{figure}[ht]
		\centering
	\includegraphics[width=\textwidth]{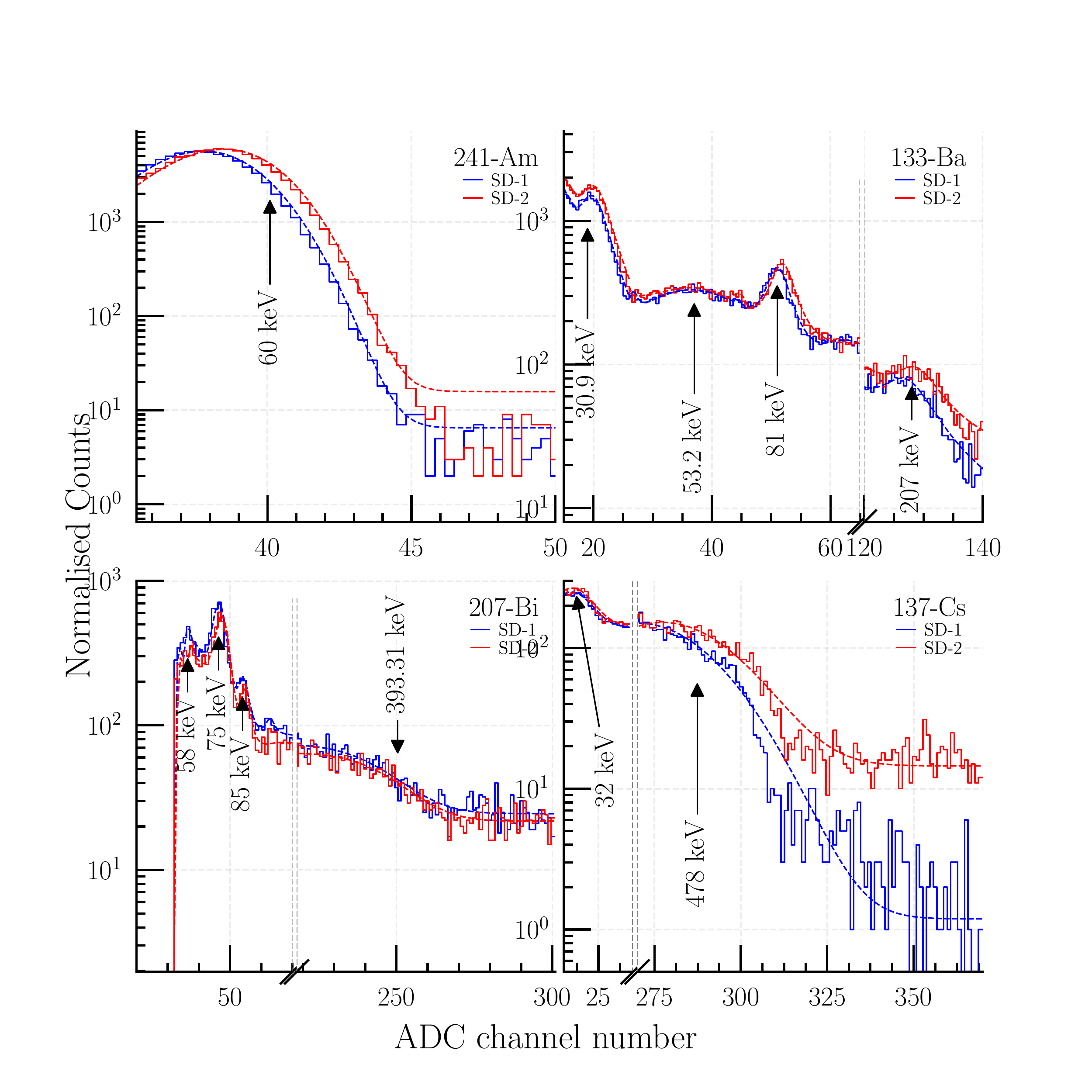}	
	{\caption[]{\label{Fig:Calibration}{\textnormal{The pulse height spectra of the calibration sources as measured by the two SDs of the prototype dosimeter.}}}}
\end{figure}

\begin{figure}[ht]
		\centering
	\includegraphics[width=14.5cm,height=10cm]{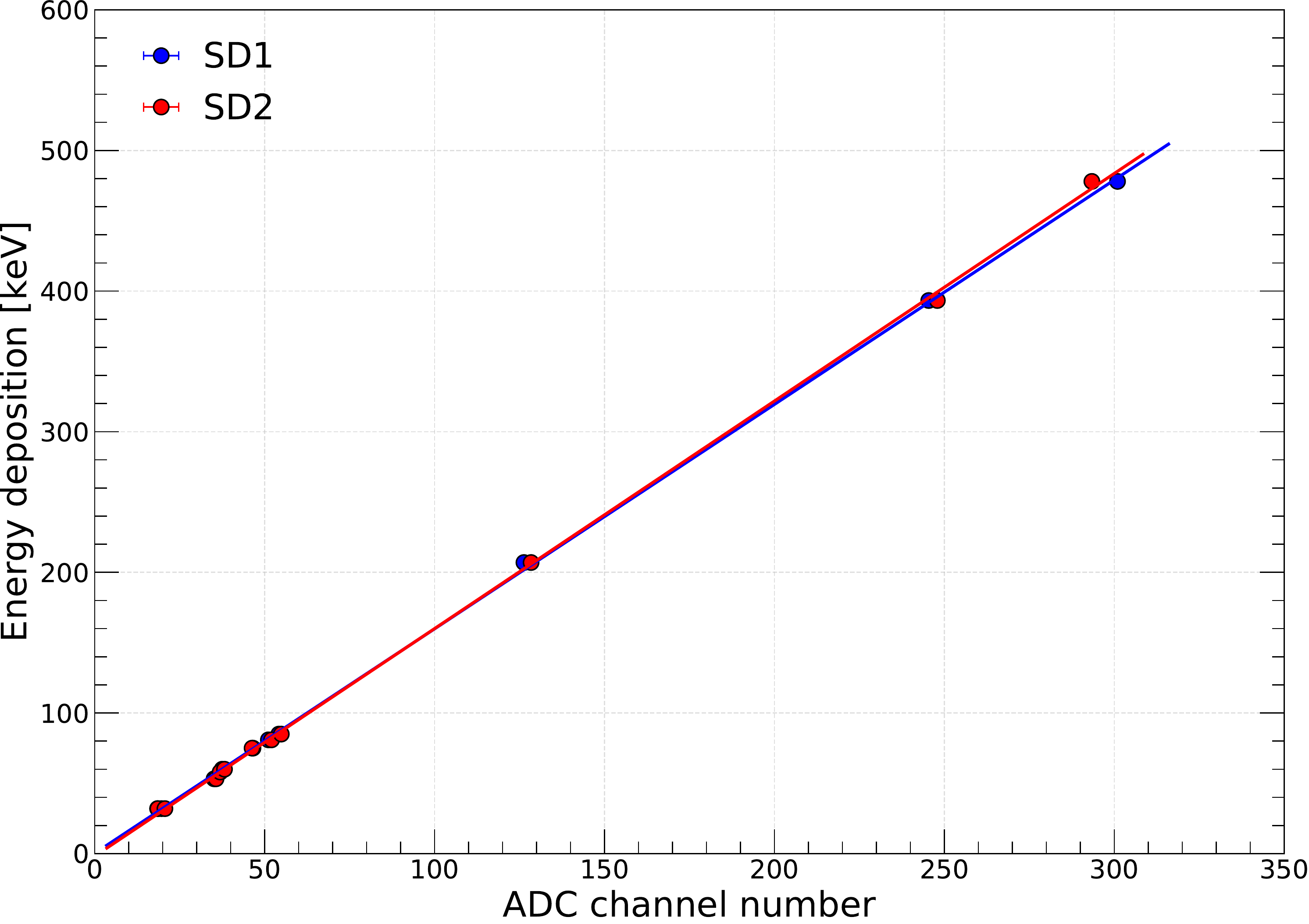}	
	{\caption[]{\label{Fig:Calibration2}{\textnormal{The resulting energy calibration curves for the high-gain channel of both SDs.}}}}
\end{figure}

We follow a similar calibration process as presented by \citet{Ritter_thesis2013} and \citet{Stephan_Bottcher2018}. In order to calibrate the RPiRENA ADC output we measure the energy spectra for the following reference sources: Americium-241 ($^{241}$Am), Caesium-137 ($^{137}$Cs), Barium-133 ($^{133}$Ba), and Bismuth-207 ($^{207}$Bi). The obtained spectra, still in terms of the ADC channel number, is shown in Figure \ref{Fig:Calibration} as binned histograms. For each source we identify a number of known characteristic photo-peaks and Compton edges (indicated by the vertical arrows). The pulse height spectra for $^{241}$Am were fitting by a combined exponential curve and Gaussian function and shown in as the dashed lines. $^{137}$Cs, $^{133}$Ba, and $^{207}$Bi also show a Compton edge and this portion of the distribution was fitted with an additional error function. The value of the Compton edge is also used in the calibrations as shown by \citet{Wuest_etal2007} and \citet{Stephan_Bottcher2018}. The calibration results are listed in Table~\ref{table:Calibration}.\\

\begin{table*}[]
		\centering
{\caption[]{\label{table:Calibration}{\textnormal{Calibration values and statistical comparisons for both SDs.}}}}
	\centering 
	\begin{tabular}{cclclc} 
		\hline\hline 
		\textbf{Type} & \textbf{Radioactive} & \textbf{Energy} & \multicolumn{2}{c}{\textbf{ADC Channels}} \\
		&  \textbf{source} & \textbf{(keV)} & \textbf{SD 1} & \textbf{SD 2} \\ 
		\hline
		Photo-peak & {$^{241}$Am-$\gamma$} & \textnormal{60.0} & 37.52 $\pm$0.04 & 38.19$\pm$0.04\\ 
		Photo-peak & {$^{133}$Ba-$\gamma$} & \textnormal{80.9} & 51.11$\pm$0.14 & 52.04$\pm$0.15\\ 
		Photo-peak &  {$^{133}$Ba-(X-ray)} & \textnormal{32.0} & 18.50$\pm$0.47 & 18.58$\pm$ 0.51\\ 
		Photo-peak &  {$^{133}$Ba-(X-ray)} & \textnormal{53.2} & 35.13$\pm$0.42 & 35.7$\pm$ 0.41\\
		Compton edge &  {$^{133}$Ba-(X-ray)} & \textnormal{207} & 126.3$\pm$0.38 & 128.4$\pm$ 0.37\\ 
		Photo-peak & {$^{137}$Cs-(X-ray)} & \textnormal{32.0} & 19.65 $\pm$0.52 & 20.66$\pm$0.40\\
		Compton edge &  {$^{137}$Cs-(X-ray)} & \textnormal{478} & 293.41 $\pm$0.33 & 300.97$\pm$0.38\\ 
		Photo-peak & {$^{207}$Bi-(X-ray)} & \textnormal{58.0} & 36.98 $\pm$ 0.29 & 36.91 $\pm$0.30\\ 
		Photo-peak &  {$^{207}$Bi-(X-ray)} & \textnormal{75.0} & 46.57 $\pm$ 0.28 & 46.31 $\pm$0.29\\ 
		Photo-peak &  {$^{207}$Bi-$\gamma$} & \textnormal{85.0} & 54.19 $\pm$ 0.17 & 54.91 $\pm$ 0.18\\
		Compton edge &  {$^{207}$Bi-$\gamma$} & \textnormal{393.3} & 245.41 $\pm$ 0.35 & 247.95 $\pm$ 0.37\\ 
		\hline 
	\end{tabular}
\end{table*}

The energy calibration points are fitted using a linear regression of the form $E=(\textnormal{A} \pm a_0)\cdot \textnormal{ADC} + (\textnormal{B} \pm b_0)$, where $E$ is the known energy deposition of the calibration sources and $\textnormal{ADC}$ the corresponding ADC channel number. The fit coefficients $\rm{A}$, $a_0$, $\rm{B}$ and $b_0$ are listed in Table~\ref{table:FItst}. The results are shown in Figure~\ref{Fig:Calibration2}. The high-gain amplifying stage used here covers energy depositions from as low as the 32 keV and, in principle, this calibration curve can be extrapolated up to 7 MeV in silicon, based on the previous works using the same low energy calibration sources by \cite{Ritter_thesis2013}. The SD 1 and SD 2 regressions shows slight difference between the SDs, ranging between 0.2\% to 5\%. This is probably due to small uncertainties in the electronics components and manufacturing tolerances \\

\begin{table}[]
	\centering
{\caption[]{\label{table:FItst}{\textnormal{Linear regression coefficients for both SDs.}}}}
	\begin{tabular}{lcccc} 	\hline\hline 
		SDs& $\rm{A}$ & {$\pm a_0$} & $\rm{B}$ & {$\pm b_0$} \\
		\hline
		SD 1 & 1.596 & 0.008 & 0.240 & 1.042\\
		SD 2 & 1.618 & 0.011 & -1.829 & 1.370\\
		\hline
	\end{tabular}
\end{table}

\subsection{Silicon diodes and their geometrical factors}
\label{subsec:geometrical}

\begin{figure}[ht]
	\centering
	\includegraphics[trim=20mm 78mm 20mm 40mm,width=0.75\textwidth]{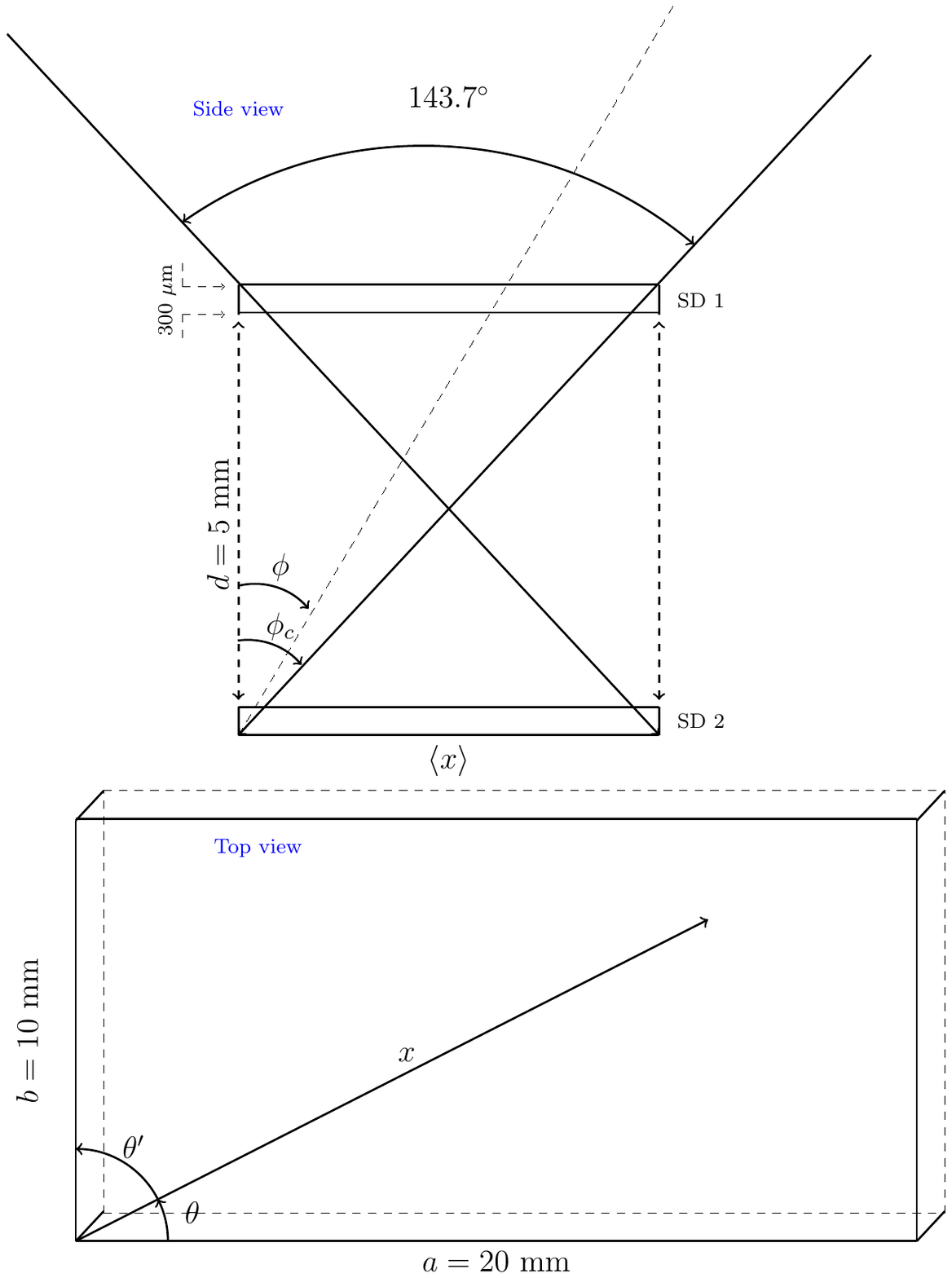}
{\caption[]{\label{fig:GF_det}{\textnormal{The s2744-08 Hamamatsu diodes in telescope mode. The top panel shows a side-on view and the SD's orientation, while the lower panel shows a top view of the SD 1.}}}}
\end{figure}

This prototype uses two commercially off-the-shelf rectangular PIN SDs (specifically model s2744-08\footnote{\url{https://www.hamamatsu.com/resources/pdf/ssd/s2744-08_etc_kpin1049e.pdf}\label{F1AppC}}) manufactured by Hamamatsu Photonics, denoted as SD 1 and SD 2 in Figure \ref{fig:GF_det} (see side-on view). The width, length, and thickness of each $\mu$SD is about 1 cm, 2 cm, and 300 $\mu$m, respectively. For ionizing radiation, however, the sensitive effective area of each diode is larger than the quoted active area for photons since energetic particles can penetrate the aluminium housing partially shielding the diodes. For dosimetric purposes the sensitive area is therefore 2.1 cm $\times$ 1.1 cm $=$ 2.31 cm$^2$ \citep{Labrenz_2014}. The SDs are arranged in a telescope configuration (with the SD 1 facing down, denoted by $\Downarrow$, and SD 2 facing up and denoted by $\Uparrow$) separated by a distance of 5 mm.\\

\subsubsection{Single diode geometric factor} 

Following the study of \citet{SULLIVAN19715}, the geometry factor (GF) is defined as the product of the effective surface area and the angular coverage of a particle detector. Assuming that the incident particles are isotropic and are interacting with the SD from one side, according to \citet{SULLIVAN19715}, the GF of a single SD is simply given by\\

\begin{equation}
{\rm GF}=\mathcal{A}\cdot2\pi\int_{0}^{\frac{\pi}{2}}\sin(\theta_2)\cos(\theta_2)d\theta_2=\mathcal{A}\cdot\pi, 
\end{equation}

where $\theta_2$ is the angle of incidence and $\mathcal{A}$ is the active effective area of a target material. Therefore, for ionising radiation interacting with the SD used in this work, ${\rm GF}=a{\cdot}b{\cdot}\pi=\textnormal{13.197}{\rm cm^2\cdot{sr}}$.\\

\subsubsection{Telescope mode geometric factor}

For these identical rectangular diodes, separated by 5 mm in telescope mode, the resulting GF is calculated from Equation 11 of \citet{SULLIVAN19715}, as $\textnormal{ 7.257}{\rm cm^2\cdot{sr}}$. \\

\subsection{Effective diode thickness}
\label{subsec:thickness}

When isotropic particles interact with both diodes, their trajectory can be projected onto the top diode, as shown in the top right panel of Figure~\ref{fig:GF_det}. We are interested in finding the maximum length of this projected tract for every angle $0^{\circ} \leq \theta \leq 90^{\circ}$. The longest possible tract is a diagonal line at $\theta_c={26.57}^{\circ}$ for these specific diodes. We now divide $\theta$ into $n$ equally spaced increments and calculate the average length of this maximum tract as

\begin{equation}
  \langle {x} \rangle = \frac{1}{n} \sum_{i}
    \begin{cases}
      \frac{a} {\cos (\theta_i)} & \text{if $\theta_i < \theta_c$}\\
       \frac{b} {\cos(\theta_i')} & \text{if $\theta_i > \theta_c$}
    \end{cases}       
\end{equation}

where $\theta'={90}^{\circ}-\theta$ and $i$ labels all the individual angles considered. For the SDs used in this work, an average maximum projected tract length of $\langle{x}\rangle =$ 15.33 mm is found.\\

We now switch to a side-on view of the particle telescope as illustrated in the top panel of Figure~\ref{fig:GF_det} and use $\langle x \rangle$ as the effective width of the detector. The maximum opening angle of the detector can now be calculated as $2 \phi_c = 143.86^{\circ}$. The effective thickness of the top diode can now be estimated if we assume an isotropic incident particle distribution from above. For particles entering along the dashed trajectory with an incident angle of $\phi$, the top diode will have an effective thickness of $l_{\mathrm{eff}} = 300\mu \textrm{m} / \cos \phi$. The average incident angle of the radiation, for an isotropic distribution, is $\langle \phi \rangle = \phi_c/2$, so that the average effective length becomes 

\begin{equation}
 \langle l_{\mathrm{eff}} \rangle = \frac{ 300\mu \textrm{m}} {\cos \langle \phi\rangle},
\end{equation}

leading to $\langle l_{\mathrm{eff}} \rangle = 370.67$ $\mu$m.\\

\section{The CARI model :  Simulations procedure}
\label{Append:B}

{The CARI model's user guide is published by Kyle Copeland\footnote{\url{https://rosap.ntl.bts.gov/view/dot/57225}}. For the two flights discussed in Sec.~\ref{Sec:Results}, simulations were performed using CARI-7A (ACADEMIC USE, version 4.1.0, published date : October 4, 2019). At aviation altitudes, neutrons and protons dominate the radiation field, while gammas, electrons, positrons, muons and pions contribute less. Once the CARI model program is started, a command interaction interface will popup. To calculate each particle contribution using the CARI-7A model, a simplistic demonstration is given below. For example, to calculated the dose rate and accumulated doses from the neutron component, we followed these options:
\begin{tcolorbox}[colback=white,sharp corners]
\begin{enumerate} 
\item[] {\bf MAIN MENU ~~~$\Longrightarrow$ option $\langle$2$\rangle$  Galactic radiation received on flights}
\item[] {\bf FLIGHT MENU $\Longrightarrow$ option $\langle$2$\rangle$  Flight path(s) defined by many waypoints (*.DEG files, same as CARI-6M and -6PM) }
\item[] {\bf Enter a file number $\Longrightarrow$ option	$\langle$1$\rangle$    JF.10.06.2019Neutron.DEG ...file will be imported }
\item[] {\bf SELECT RADIATION  $\Longrightarrow$ option	$\langle$1$\rangle$  ...for neutron doses}
\item[] {\bf SELECT COSMIC RAY MODEL $\Longrightarrow$	option $\langle$1$\rangle$   GCR: ISO TS15390:2004-MSU-NYMMIK }
\item[] {\bf SELECT DOSE TYPE $\Longrightarrow$	option $\langle$7$\rangle$  ABSORBED DOSE IN 0.3 mm Si chip }
\item[] {\bf USE THE SUPERPOSITION APPROXIMATION $\Longrightarrow$	option $\langle$Y,N$\rangle$  Y }
\item[] {\bf USE DATES FROM FLIGHT DATA FILE $\Longrightarrow$	option $\langle$Y,N$\rangle$  Y }
\end{enumerate}
\end{tcolorbox}
}
 \begin{figure}[ht]
	\centering
	\includegraphics[trim=10mm 6mm 10mm 8mm,width=0.9\textwidth]{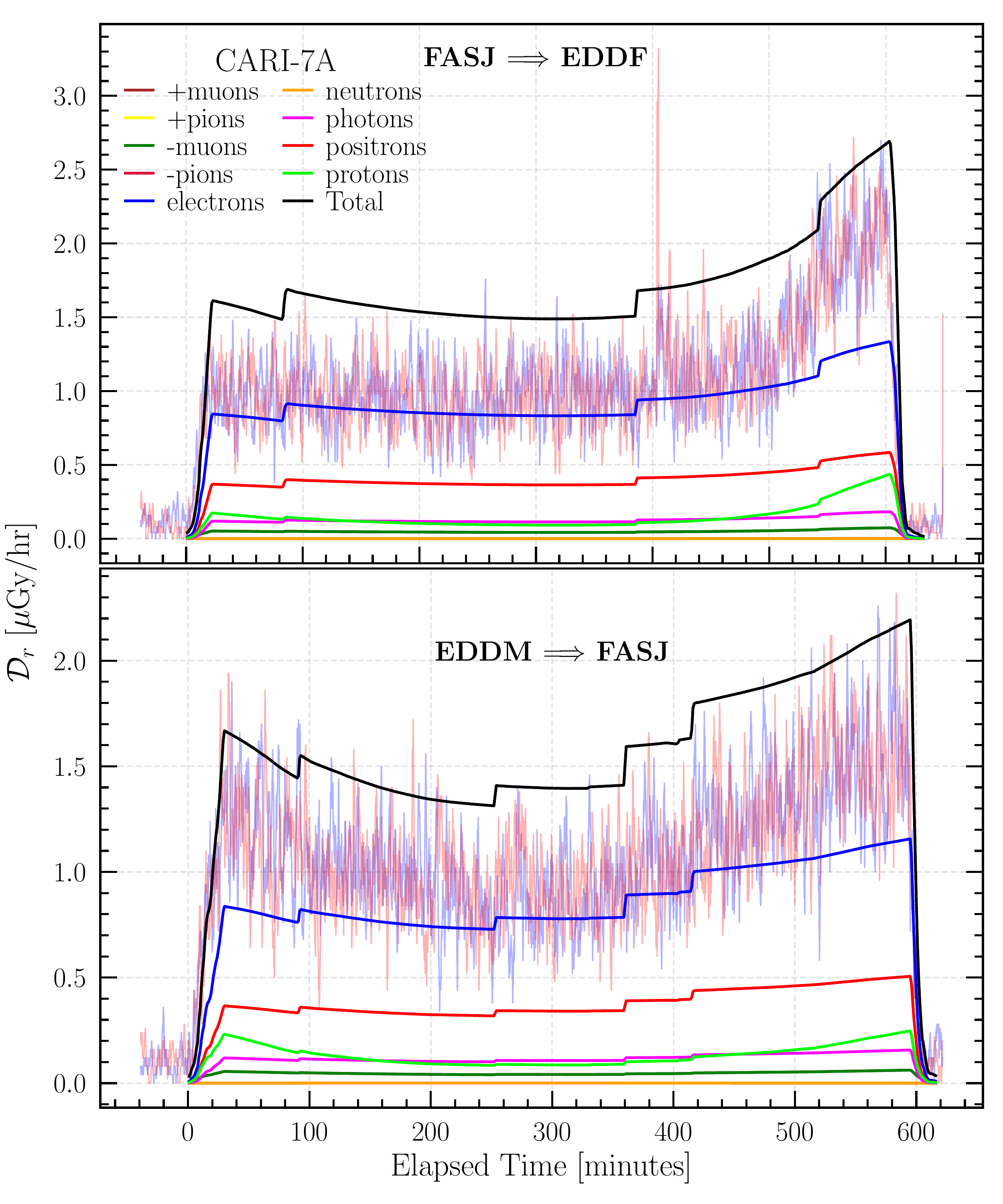}
	\caption[]{\label{fig:CARI7}{{\textnormal{{ Comparison of the CARI-7A model calculated absorbed doses in a 0.3 mm Si diode to the RPiRENA mesured doses. The red and blue fluctuating lines represents the SD1 and SD2 RPiRENA diode measurements, respectively.}}}}}
\end{figure}
 
{Figure \ref{fig:CARI7} shows the calculated dose contributions from different radiation sources (colour coded) and the total dose contribution (i.e. the sum of all the individual doses; the black solid line). These calculated absorbed doses in a 0.3 mm Si diode are compared to the RPiRENA mesured doses (see the blue and red solid lines representing the SD1 and SD2 measurements, respectively).}

\cleardoublepage

\bibliographystyle{swsc}
\bibliography{biblio}

\end{document}